\documentstyle[12pt,epsf,aaspp4]{article}
\begin{document}
\slugcomment{Submitted to ApJ 4/30/96}
%
%
\title{Strong Clustering in the Low Redshift Lyman-$\alpha$ Forest}
\author{Andrew Ulmer\altaffilmark{1}}
\affil{Princeton University Observatory, 
       Princeton, NJ~08544-1001}
\altaffiltext{1}{andrew@astro.princeton.edu}
%
%
\begin{abstract}

The two-point correlation function, $\xi$, of Lyman-$\alpha$ forest
is found to be large, $\xi = 1.8^{+1.6}_{-1.2}$, $>90\%$
confidence level, on the scale of 250-500 km/s
for a sample of absorbers $(0 < z < 1.3)$ assembled from
HST Key Project Observations.
This correlation function is stronger than
at high redshift ($z > 1.7$) where $\xi \approx 0.2$ for
velocities $> 250$~km/s.

\end{abstract}

%
%
\section{Introduction}

Many correlation studies of the Lyman-$\alpha$ forest have been made
in the past decades.
Almost all of these investigations were limited to
redshifts greater than $\sim$1.7, since they were carried out on the ground.
After many years of controversial weak detections of the two-point
correlation function at $z> 1.7$ on scales $<300$~km/s
(\cite{web87,ost88,rau92,sri94,che95,crie95,kul96,sri96}),
a consensus may be developing that there is a small,
but finite correlation, $\xi \approx 0.2$
on scales less than 300 km/s (see \cite{cri95}~ for a review).
In contrast, based on the finding of Cowie~et~al. (1995), which shows that
some high-redshift Ly-$\alpha$ lines have many weak associated carbon lines
described as ``cloudlets,'' Fern{\'{a}}dez-Soto et~al.~ (1996)
report an extremely strong correlation function, $\xi \approx 10$,
among these weak carbon lines on scales $\sim 200 km/s$.
This finding begs the question of what an individual Ly-$\alpha$ system is.
At scales larger than 300 km/s, the correlation at $z > 1.7$ is
to quite weak or undetectable (\cite{sar80,sar82,bec87,web91,cri95,sri96}).

Recently, UV observations with Faint Object Spectrograph
of the Hubble Space Telescope have opened up a new regime of
redshift space for study, $z\approx$~0-1.3
(Bahcall et~al. 1991, 1992, 1993a (CAT1), 1993b, 1996 (CAT2);
Morris et~al.~ 1991).
A reasonably strong correlation function among these low-redshift systems
might be anticipated because (1) gravitational correlations tend
to grow with time, and (2) very low redshift observations of galaxies
near quasar lines of sight which show that perhaps one third or more
of the Ly-$\alpha$ forest lines are associated with galaxies
(\cite{lan95}; see also \cite{bah92,spr93,sar95}) which
are highly correlated objects.
No evidence for a two-point correlation function was found in the first
analysis of the Quasar Absorption Line Key Project data (CAT1),
although evidence for clumping of
the Ly-$\alpha$ forest lines near metal-line systems was discovered (CAT2).
Studies of high redshift metal-line systems
show strong two-point correlation functions
($\xi \stackrel{>}{\sim} 5$ at 100 km/s) on scales of
up to $\sim$500~km/s (\cite{sar88,hei89,pet94}).
Therefore, if the low-redshift Ly-$\alpha$ forest lines are either
gravitationally evolved counterparts of high redshift Ly-$\alpha$ lines,
are associated with ordinary galaxies, or clump around the strongly
clustered metal lines, they should have a relatively
strong, yet unmeasured, two-point correlation function.

In this paper, a complete, well-defined sample of low-redshift absorbers
is analyzed for a two-point correlation function using techniques that
are especially appropriate for quasar absorption lines.
In particular, a new algorithm for simulating data with a two-point
correlation function allows for a good determination of the
errors in the detected correlation function.
The paper is organized as follows.
In \S II, the data are described and the procedure for creating a
complete data set is presented.
Using extensive Monte Carlo simulations, the two-point correlation
function for the Ly-$\alpha$ forest is shown to be non-zero in \S III.
In \S IV, the strength of the two-point correlation is determined, again
with use of Monte Carlo simulations. 
Blends are shown to have only a minimal effect on the findings in
\S V.
A discussion and summary of the results of the paper is given in
\S VI.

\section{A Complete Data Set}

In order to create a sample free of experimental biases,
a subset of Ly-$\alpha$ lines was selected
from the published catalogs assembled by the
Quasar Absorption Line Key Project from Faint Object Spectrograph (R=1300)
observations on the Hubble Space Telescope (\cite{cat1,cat2}).
The selection procedure is similar to that
described in \S VII of \cite{cat2}. 
The Ly-$\alpha$ sample adopted here has a uniform detection limit of 0.24~\AA,
rest equivalent width, 
which corresponds to $\sim 2 \times 10^{14}$~HI/cm$^2$ for unsaturated lines.
Lines from spectral regions in which the line detection limit,
$W_{\rm{min}}$, was greater than 0.24~\AA~ are excluded.
In other words, no lines are included in the sample
which are very weak lines or which come from parts of the
spectrum where the sensitivity was low.
Because of the proximity effect, lines within 3000 km/s of the
quasar emission redshift are excluded.

The total sample consists of 100 Ly-$\alpha$ lines. Nine of these
are associated with  extensive metal-line systems with four or more detected
metal lines (see \cite{cat1,cat2}).
The mean redshift of the lines is 0.7.

The complete sample used for analysis in this paper
is given in Table~\ref{cat1tab} and Table~\ref{cat2tab}.
The spectral range for each quasar spectrum where the sensitivity
was high enough to meet the uniform detection limit criteria is given
in Table~\ref{covtab}.

\section{Pairs counts and the Two-point Correlation Function}

The two-point correlation function can be estimated
from the pair counts in the data according to the formula,
\begin{equation}
\label{xi}
\xi(r)= \frac{N({\rm Obs})(r)}{N({\rm Ran})(r)}-1, 
\end{equation}
where $N({\rm Obs})$ is the number of observed pairs and $N({\rm Ran})$ is
the number of pairs in a given redshift bin
that would be expected in the absence of clustering.
$N({\rm Ran})$ is determined by Monte Carlo simulations.
The same velocity bins are used as employed
in the first empirical analysis of the two-point correlation function
in this data (\cite{cat1}).

In the Monte Carlo simulations,
the number of pairs expected in the absence of clustering
was calculated by placing lines along
the part of each quasar line-of-sight where the minimum observable
rest equivalent width was less than the uniform detection limit of 0.24~\AA.
The lines are Poisson distributed with the
probability of a line being placed at a given redshift being
proportional to the assumed density at the corresponding redshift.
Studies of the number density at both high and low redshift have indicated that
\begin{equation}
\label{numden}
n(z) \approx A (1+z)^\gamma,
\end{equation}
where $n(z)$ is the number density, $A$ is a normalization constant, and
the exponent $\gamma$ is between 0 and 1.5 in the redshift range
of interest (CAT1,CAT2).
To mimic the finite resolution of the
data in the simulations,
lines with separations less than 230 km/s were excluded
following \cite{cat1}.
For each separate data subset investigated, ten thousand simulations
were performed which was adequate to fix the error bars reliably.
The slope, $\gamma$, of the number density was varied between
0 and 1.5, which covers the range allowed by the observational data.

The approach taken here differs from some of
the previous analyses of clustering (e.g. \cite{cat1}) in two ways.
First, the normalization constant, $A$, used to produce the
number of lines along each line-of-sight, is determined for the entire
sample and is the same for every quasar line-of-sight.
In previous papers, the Monte Carlo simulations, used to determine
$N({\rm Ran})$ matched the number of simulated lines along the line-of-sight
with the observed number of lines along the line-of-sight.
The strength of the correlation function will be underestimated
if $A$ is determined, in this way, separately, for each line-of-sight.
Monte Carlo simulations show the strength of this normalization
effect in this data sample to be about 20\% in the sense that
the expected number of pairs is decreased by 20\% when the normalization
done correctly.
The effect is most important in data samples with a small
number of lines along each line-of-sight where Poisson fluctuations of
the number of lines along each line of sight are significant, as is
the case for low-redshift studies.
Second, the evolution of number density evolution is included in
the Poisson distribution.
If it were neglected, the number of expected pairs would
be underestimated, and the strength of any clustering would be overestimated,
because pair counts scale as the square of the density,
\begin{equation}
\Delta N({\rm Ran}) = N({\rm Ran}) \left[ \frac{\int n^2(z) {\rm d}z} 
{\int n^2 {\rm d}z} -1 \right],
\end{equation}
where $n(z)$ is the number density as a function of redshift, and
$n$ is the unevolving number density.
In this sample, the strength of this effect is about 20\% as well.

Figures~1~and~2 show the pair counts of Ly-$\alpha$ systems alone
and for the combined Ly-$\alpha$ and metal-line system subset.
For Figure~1, pair separations are calculated in terms of
relative velocity as measured by an observer at the absorption
systems, and for Figure~2 the separations are shown in terms of
comoving coordinates.
The relations used to calculate the coordinates are ($\Omega =1$)
\begin{equation}
\Delta v =  \frac{c~ \Delta z}{1+z},
\end{equation}
and
\begin{equation}
\Delta r =  \frac{c~ \Delta z}{H_0(1+z)^{1.5}}.
\end{equation}
There do not appear to be any statistically significant differences when
the data are analyzed in terms of comoving distance rather than
velocity.
The slope of the number density affects the expected number of
pairs in the absence of clustering, $N({\rm Ran})$, and therefore
the strength of any correlation.
The largest number of expected pairs is found with the largest value of
$\gamma$ (1.5) and the smallest number of expected pairs with
$\gamma = 0$.
The results of the Monte Carlo simulations
for ($\gamma =  0 ~{\rm and}~ 1.5$) are also shown in Figures~1~and~2.
The actual slope of the number density is likely between these
values, and correspondingly, the true $N({\rm Ran})$ lies between the
values resulting from these Monte Carlo simulations.

For both samples and separation measures, there is a
strong excess of pairs in the first bin corresponding to
250-500 km/s or 2.5-5 $h^{-1}$Mpc.
Table~\ref{probtab} shows the results of the Monte Carlo experiments that were
designed to determine the statistical significance of the observed
correlation.
The probability that an unclustered distribution could produce in
the first bin an excess
as high as observed is less than 0.5\% in all the cases considered.
The probabilities reported in Table~\ref{probtab} are conservative in the
sense that, $\gamma$, was taken to be 1.5
which is as large as allowed by observation at $2 \sigma$, so that
$N({\rm Ran})$ was a large as possible.

\section{Strength of the Correlation Function}

The allowed range of strength of the two point correlation function
is determined by Monte Carlo simulations (see Table~\ref{probtab}).
Because the correlation function is detected in only the first bin of the
pair counts, the shape of the correlation function was taken as
a free parameter.
As shown below, this freedom does not strongly
affect the range of the determined strength.

To test the strength of the observed correlation function by Monte Carlo
simulation, an algorithm was devised to create
a one-dimensional sample with a specified two point correlation function.
The many 3-dimensional methods for generating correlated sets of objects
(e.g. \cite{ney52,son77}) diverge in 1-dimension.
Therefore, a different algorithm was developed for the 1-dimensional
application in this paper.
Details of the algorithm and its relation to 3-D methods
will be described in detail in a future paper (\cite{ulm96});
a brief description is supplied in Appendix A.

Confidence regions were established by
determining what the strongest (weakest) simulated correlation function
for which X\% of the simulations had as few (as many) pairs in the
first bin as the data.
Evolution of the number density (Eq.~\ref{numden})
has an effect on the expected number of pairs in the absence of
clustering, and therefore on the strength of the correlation function
(see Figures~1~and~2).
Table~\ref{probtab} lists the minimum, $\xi_{\rm min}$, and the maximum,
$\xi_{\rm min}$, values of the correlation function evaluated
at the 90\% confidence limits in the 250-500 km/s bin.
The value of $\xi_{\rm min}$ ($\xi_{\rm max}$) was established by requiring
that less than 5
functions smaller (larger) that the value found in the actual sample.
The confidence regions given in Table~\ref{probtab} account
for this uncertainty in a conservative fashion.
For instance, the upper limit of the confidence region, $\xi_{\rm max}$, is
found using simulations with the smallest allowable number density slope,
$\gamma = 0$, which implies the smallest number of expected pairs, and
therefore the largest correlation function.
Then,  $\xi_{\rm max}$, is equal to the value of the correlation
function, evaluated in the 250-500 km/s bin, which has the number of pairs
in the first bin larger than those observed in 95\% of the trials.

In principle, the confidence regions will depend on all n-point correlation
functions, or equivalently, the exact algorithm used to create the simulation.
In this particular case where the pair counts in only one bin are examined,
the 90\% confidence regions given in Table~\ref{probtab}
is not expected to be too sensitive to the particular type of correlation
function used in the simulated data.
To test this hypothesis, a variety of two-point
and three-point correlation functions were simulated and thereby
used to construct confidence intervals.
The intervals derived were found to be consistent with those reported in
Table~\ref{probtab} to within 5\% of their values.
Specifically, confidence regions were determined from simulated data with
two-point correlation functions of the form which describes
galaxies and clusters  (\cite{peb80})
\begin{equation}
\begin{array}{ll}
\xi(r) = -1; & (r < r_{\rm min}) \\
\xi(r) = (r/r_0)^{-\gamma}; & (r > r_{\rm min}) 
\end{array}
\end{equation}
with $r_{\rm min}$ corresponding to the resolution of the data and the
exponent, $\gamma$, ranging between 1.3 and 2.3.
A particular type of three-point correlation function which is
appropriate for galaxies was also added
with a strength parameter Q ranging from 0.5-1 (\cite{peb80}).
While the Ly-$\alpha$ forest may not have the
same two and three point correlation functions as described above,
the important point is that the confidence regions of the
strength of the correlation function are insensitive to the
specific shape of correlation function.

The top panel of Figure~3 illustrates the confidence intervals for
one of the cases in Table~\ref{probtab}.
For comparison, the galaxy-galaxy correlation
function is shown in the bottom panel of the figure.
The galaxy-galaxy correlation function is within the
confidence limits determined by Monte Carlo simulations.

\section{Influence of Blends}

In this section, the reality of the close pairs and the possible
influence of blends on the observed pairs is investigated.
Table~\ref{pairtab}
gives detailed information about each pair in the 250-500 km/s
bin which shows the excess.
The average and median equivalent widths of the lines in the
close pairs are indistinguishable from that of the entire sample.

In the absence of clustering,
the low number density of Ly-$\alpha$ lines at small redshift
makes blending much less of a problem than at larger redshifts.
The resolution of the Faint Object Spectrograph is
1300 (CAT1) which corresponds to 230 km/s.
Then, the number of expected blended lines, that is, lines with
separations less than 230 km/s, in the entire sample
is $\sim$4.
If the blended lines
were separated by the line selection algorithm into pairs of
lines with separations greater than 250 km/s, an artificial excess
of close pairs could be created in the sense that the
250-500 km/s bin would contain all pairs with actual separations
between 0 and 500 km/s.
This contamination is
unlikely to contribute strongly to the excess detected pairs, because
such artificial pairs should have separations just larger than
the resolution limit, whereas the close pairs shown in
Table~\ref{pairtab} do not show this property.

A single line convolved with the instrumental profile and
sufficient noise
may, in principle, be artificially split into a pair by the automated
line selection algorithm.
The efficiency of such splittings can be assessed
by Monte Carlo experiments using
the detailed simulation software developed by the QSO absorption Key Project
team (\cite{sch93}).
D. Schneider (private communication, 1996)
generously ran 6000 lines through
the Key Project line-simulation software which adds noise, accounts for
instrumental resolution, and implements the same line selection routine
that was used on the actual Hubble data. The results are very encouraging.
Even, if all 100 lines in the sample had a huge internal velocity
dispersion of 300 km/s (typical velocity dispersions are 15-50 km/s at
$z\sim 2$, e.g. \cite{kul96}),
and a large rest equivalent width of 0.67 as did the first 3000 simulated
lines
(the actual lines are distributed as an exponential distribution with
scale $\sim$0.32~\AA),
the estimated total number of lines expected to be artificially
split into lines with separation greater than 250 km/s
in the entire data set is only 0.56.
The actual number of artificial pairs in the data is probably much smaller.
First, many of the lines in the sample are so weak that even they
could not be split into two lines which would exceed the uniform detection
limit of 0.24\AA, rest equivalent width.
The effective cutoff in rest equivalent width for a line
to be be able to split into two lines in the sample occurs above
$2\times 0.24 =0.48$\AA, and is closer to 0.6\AA, because lines do not
often split exactly into equal parts.
The fraction of lines in the sample with rest equivalent width $> 0.6$\AA~
is $\sim0.3$, so the total number of expected artificial pairs is reduced to
$\sim$0.17.
Further, most Ly-$\alpha$ lines probably have velocity dispersions much
less than 300 km/s, and are therefore less likely to split into
artificial pairs.
We conclude that the effect of blends is no more than 5\% on the
total number of detected pairs and probably substantially less.
Even in the pathological case that all lines had gigantic velocity
dispersions of 500 km/s which was the case for the second set of 3000
simulated lines, simulations and the same line of argument shows that
the total number of artificial pairs is expected to
be less than 0.7.
This pathological case would affect slightly the confidence regions described
in the previous section, but would still not account
for all of the excess close pairs.

Because the close pairs are real, the Ly-$\alpha$ lines are strongly
clustered. This clustering will effect the number of
blends in the medium resolution data and has
important implications for further analyses of the Ly-$\alpha$
forest in this medium resolution data set.
For instance, correlations
affect the answer to the question: if the data could be viewed
under higher resolution, would many of the lines be resolved
into multiple systems?
Under the hypothesis that all lines are unclustered, one expects only about
4 (of 100) lines to be composed of
two smaller lines (by random superpositions), 
because of the extremely low line-of-sight density of absorbers.
The difference in inferred volumed density of of Ly-$\alpha$ absorbers
would be only $\sim 4$\% as obtained from
high and low resolution observations.
However, the lines appear to be strongly correlated, and the number of
expected unresolved pairs (with $\Delta v < 230$~km/s)
is at least as large the total number of resolved pairs
with 230~km/s~$< \Delta v <$~460~km/s -- about 10.
If the correlation function increases for velocities less than 250 km/s,
as may be the case for high redshift, higher resolution studies,
then the number of unresolved pairs may be more than twice as large, or 20
(this would be the case, for instance, if the shape of the correlation
function were a Gaussian with zero mean and scale $\sim 250$km/s).
In the absence of clustering, resolution is expected to make a very small
difference in the determination of the statistics of the Ly-$\alpha$ forest,
e.g. the number density and number density evolution.
Viewed under high resolution and with strong clustering,
the number density and evolution of Ly-$\alpha$ forest may easily differ
by factors of $\sim$20\% from their presently determined values.

\section{Discussion and Conclusion \label{discussion}}

The two-point correlation function, $\xi$, of Ly-$\alpha$ forest lines,
$\left<z\right> = 0.7$, is $\xi = 1.8^{+1.6}_{-1.2}$, 90\% confidence level,
for separations of 250-500 km/s (see Table~\ref{probtab}).

How does the clustering at low redshifts
compare with the recent detections of high redshift clustering?
Between redshifts of 1.7 and 4, for similar velocity separations
and equivalent width limits,
it seems likely that the correlation function is 
$\stackrel{<}{\sim}0.2$ on scales greater than 250 km/s (see \cite{cri95}
and references therein; see \cite{fer96} for a different view).
Additionally, Cristiani's investigation
shows some evidence for the increase of the correlation function
(for velocity separations of 100 km/s) with decreasing redshift.
Similar findings have also been reported by Srianand (1996). 

In velocity space, the correlation function has strengthened
with redshift from $\xi \stackrel{<}{\sim}0.2$ at high redshifts to
$\xi \stackrel{>}{\sim} 1$ at the small redshifts considered.
The interpretation of the increasing correlation function
is not straightforward.
For instance, the low and high redshift absorbers
may not be the same (see the discussion in \cite{cat2}
and references therein).

How does the detected strength of
the Ly-$\alpha$ correlation function compare to that of galaxies?
Such a comparison is motivated in part by the long-standing idea that
Ly-$\alpha$ absorbers are related to the halos of galaxies (e.g.
\cite{bah79,mo94})
as well as by the finding that at very low redshift, perhaps one third
or more of the Ly-$\alpha$ lines are associated with galaxies (\cite{lan95}).
A zeroth-order comparison is made in the second panel of Figure~3 by plotting
the galaxy-galaxy correlation function ($z=0$)
against the observed Ly-$\alpha$ correlation function.
The two correlation functions are consistent within the limits of the data.
Even the simplest galactic-halo Ly-$\alpha$ forest models have as many as 4
free parameters: velocity dispersion of galaxies as a function of redshift,
evolution of the galaxy-galaxy correlation function,
fraction of galaxies with Ly-$\alpha$ lines, and the truncation point
of the spatial two-point correlation function.
Therefore, a detailed analysis of the idea that galactic halos and Ly-$\alpha$
lines are related will have to wait until
greater amounts of data are collected and good estimates of both
the amplitude and the shape of the correlation function become available.

\acknowledgements
AU is supported by a NSF Graduate Research Fellowship. I thank J. N. Bahcall,
J. Miralda-Escude, and H. J. Mo for stimulating and informative discussions,
and D. P. Schneider for helping to investigate the possibility of
false line pairs in the observations.

\newpage

\section{Appendix A}

In this appendix, the algorithm for simulating data with a given two-point
correlation function is briefly described.
A further description as well as sample code for implementing
the algorithm will be provided elsewhere (\cite{ulm96}).
It was necessary to devise a new algorithm, because traditional
3-dimensional algorithms (e.g. \cite{ney52,son77}) diverge in 1-dimension.

The algorithm generates a 1-dimensional set of points with two-point
correlation functions of the form
\begin{equation}
\label{appendxi}
\begin{array}{ll}
\xi(r) = -1; & (r < r_{\rm min}) \\
\xi(r) = (r/r_0)^{-\gamma}; & (r > r_{\rm min}) 
\end{array}
\end{equation}
There are three free parameters in the correlation function:
$\gamma$, $r_0$ and  $r_{\rm min}$.

The algorithm can be motivated by writing the conditional probability of
finding one object in the element $dr$ at a distance $\Delta r$ from
another object,
\begin{equation}
\label{prob}
dP=n(1+\xi(\Delta r))dr,
\end{equation}
where $n$ is the density.
A first approach to generating a 1-dimensional data set with a two-point
correlation function could be to
(1) start with an object at $r=0$ and (2) place a second object at
$r_1=\Delta r_1$ according to a Poisson process with probability of an
object being placed at $\Delta r$ ($>0$) as determined by Eq.~\ref{prob},
(3) place the third object at $r_2=r_1+\Delta r_2$
according to Eq.~\ref{prob}, that is place the object only with reference to
the one directly preceding it (4) place the $i$th object at
$r_i=r_{i-1}+\Delta r_{i}$, and so on.
The simulated data produced by this algorithm has some flaws.
First, the density of the simulated data is not necessarily $n$, because the
probability of placing an object is smaller than $ndr$ when $r<r_{\rm min}$
and greater than $ndr$ when $r>r_{\rm min}$, and they need not cancel out.
Second, the slope and scale of the two-point correlation in the simulated data
are not exactly the same as specified by the input parameters,
$\gamma$, $r_0$ and  $r_{\rm min}$.

These problems can be solved by using a slightly different correlation
function $\tilde{\xi}$ with an extra parameter, $\epsilon$,
that can be tuned so that
the algorithm prodeces the correct density.
We write this correlation function as
\begin{equation}
\label{appendxitil}
\begin{array}{ll}
\tilde{\xi(r)} = -1; & (r < \tilde{r_{\rm min}}) \\
\tilde{\xi(r)} = (r/\tilde{r_0})^{-\tilde{\gamma}} -\epsilon; &
(r > \tilde{r_{\rm min}}),
\end{array}
\end{equation}
where $\epsilon$ is between 0 and 1. The other input parameters,
$\tilde{\gamma}$, $\tilde{r_0}$ and  $\tilde{r_{\rm min}}$,
differ slightly from $\gamma$, $r_0$ and  $r_{\rm min}$.
The final algorithm is therefore identical to the one give above expect
that the probability of finding a point at a given separation, $\Delta r$
from another is given by
\begin{equation}
\label{prob2}
dP=n(1+\tilde{\xi}(\Delta r))dr
\end{equation}

The algorithm is able to produce data with the desired two-point correlation
function with high accuracy to over 10 scale lengths in radius for
the densities and correlations appropriate for this study.

\newpage
\baselineskip 16pt

\begin{table}[]
\caption{ Lyman-alpha lines in sample from the first HST Key Project
catalog (CAT1).
Daggers ($^\dagger$) denote extensive metal-line systems.
Equivalent widths are given in the rest frame.
All values are given in Angstroms.
\label{cat1tab}}
\begin{tabular}{lll|lll|lll}
$\lambda_{\rm obs}$ 
& EqW &
Object  &
$\lambda_{\rm obs}$ 
& EqW &
Object  &
$\lambda_{\rm obs}$ 
& EqW &
Object  \\
\hline
1743.3   &0.33&3C 95     &1486.6   &0.34&3C 351    &2046.0   &0.43&PKS 2145+06\\
1750.2   &0.49&3C 95     &1535.0   &0.30&3C 351    &2104.1   &0.49&PKS 2145+06\\
1910.6   &0.47&3C 95     &1561.5   &0.62&3C 351    &2165.1   &0.69&PKS 2145+06\\
1959.9   &0.40&3C 263    &1563.9   &0.37&3C 351    &2177.6$^\dagger$&1.22&PKS 2145+06\\
1980.7   &0.35&3C 263    &1363.5   &0.53&H 1821+643&2210.1   &0.25&PKS 2145+06\\
1296.5   &0.30&3C 273    &1365.4   &0.32&H 1821+643&2247.7   &0.55&PKS 2145+06\\
1482.7   &0.54&PG 1259+593&1394.1  &0.58&H 1821+643&2258.3   &0.38&PKS 2145+06\\
1487.2   &0.25&PG 1259+593&1422.8  &0.64&H 1821+643&2285.1   &0.71&PKS 2145+06\\
1571.6   &0.34&PG 1259+593&1435.2  &0.30&H 1821+643&2082.6   &0.27&3C 454.3   \\
1411.5   &0.29&3C 351    &1475.4   &0.46&H 1821+643&2118.1   &0.59&3C 454.3   \\
1444.2   &0.36&3C 351    &1489.4   &0.75&H 1821+643&2211.6   &0.33&3C 454.3   \\
1485.0$^\dagger$&0.93&3C 351    &1576.9$^\dagger$&0.52&H 1821+643&2215.6   &0.47&3C 454.3   \\
\end{tabular}
\end{table}

\begin{table}[]
\caption{
Lyman-alpha lines in sample from the second HST Key Project
catalog (CAT2).
Daggers ($^\dagger$) denote extensive metal-line systems.
Equivalent widths are given in the rest frame.
All values are given in Angstroms.
\label{cat2tab}}
\begin{tabular}{lll|lll|lll}
$\lambda_{\rm obs}$
& EqW &
Object  &
$\lambda_{\rm obs}$
& EqW &
Object  &
$\lambda_{\rm obs}$
& EqW &
Object  \\
\hline

1866.0   &0.29&TON 153    &2276.6   &0.40&PKS 0122-00&2332.4   &0.32&PG 1634+706\\ 
1949.9   &0.32&TON 153    &2303.5   &0.70&PKS 0122-00&2342.3   &0.46&PG 1634+706\\ 
1983.1   &0.45&TON 153    &1843.0   &0.28&PG 1352+011&2388.7   &0.77&PG 1634+706\\ 
2018.8$^\dagger$&1.48&TON 153    &1846.5   &0.45&PG 1352+011&2412.0   &0.45&PG 1634+706\\ 
2029.2   &0.60&TON 153    &1849.0   &0.25&PG 1352+011&2416.8   &0.25&PG 1634+706\\ 
2032.1   &0.39&TON 153    &1852.0   &0.41&PG 1352+011&2420.1$^\dagger$&1.10&PG 1634+706\\ 
2034.5   &0.78&TON 153    &1855.1$^\dagger$&2.62&PG 1352+011&2435.9   &0.60&PG 1634+706\\ 
2036.9   &0.78&TON 153    &1894.3   &0.37&PG 1352+011&2453.0   &0.51&PG 1634+706\\ 
2038.8   &0.27&TON 153    &1941.7   &0.39&PG 1352+011&2481.6$^\dagger$&1.42&PG 1634+706\\ 
2060.9   &0.73&TON 153    &1946.7   &0.76&PG 1352+011&2501.4   &0.46&PG 1634+706\\ 
2107.9   &0.54&TON 153    &1989.6   &0.48&PG 1352+011&2538.5   &0.62&PG 1634+706\\ 
2118.1   &0.26&TON 153    &2019.0   &0.52&PG 1352+011&2560.9   &0.37&PG 1634+706\\ 
2159.4   &0.41&TON 153    &2022.1   &0.34&PG 1352+011&2564.9   &0.28&PG 1634+706\\ 
2178.0   &0.29&TON 153    &2027.6$^\dagger$&0.94&PG 1352+011&2599.9   &0.90&PG 1634+706\\ 
2223.9   &0.25&TON 153    &2062.3   &0.48&PG 1352+011&2603.3   &0.48&PG 1634+706\\ 
2244.5   &0.30&TON 153    &2107.2   &0.43&PG 1352+011&2628.7   &0.24&PG 1634+706\\ 
2254.5   &0.43&TON 153    &2150.8   &0.44&PG 1352+011&2672.0   &0.31&PG 1634+706\\ 
2276.4   &0.42&TON 153    &2188.2   &0.28&PG 1352+011&2677.4   &0.25&PG 1634+706\\ 
2077.9   &0.82&PKS 0122-00&2212.0   &0.32&PG 1352+011&2706.9   &0.47&PG 1634+706\\
2098.8   &0.28&PKS 0122-00&2224.5   &0.25&PG 1352+011&2778.1   &0.25&PG 1634+706\\
2226.9   &0.47&PKS 0122-00&2317.2$^\dagger$&0.49&PG 1634+706& \\
2265.7   &0.58&PKS 0122-00&2324.5   &0.67&PG 1634+706& \\
\end{tabular}
\end{table}

\begin{table}[]
\caption{
Coverage of each observed quasar spectrum (CAT1, CAT2) in which
the sensitivity to meet the uniform detection limit of 0.24~\AA~
described in \S 2 is achieved. Wavelengths are given in Angstroms.
\label{covtab}}
\begin{tabular}{lll}
$\lambda_1$ &
$\lambda_2$ 
& Object  \\
\hline
 -   &   -& 	PG 0043+039\\
1906 &1953&	PKS 0044+03\\
1728 &1941&	3C 95	   \\
1795 &1820&	US 1867		\\
1903 &1987&	3C 263		\\
1235 &1393&	3C 273		\\
1335 &1600&	PG 1259+593	\\
1641 &1648&	B2 1512+37	\\
1332 &1600&	3C 351		\\
1319 &1600&	H 1821+643	\\
2022 &2307&	PKS 2145+06	\\
2032 &2307&	3C 454.3	\\
 -   & -  &	PKS 2251+11	\\
1733 &2307&	TON 153		\\
2030 &2307&	PKS 0122-00	\\
1767 &2307&	PG 1352+011	\\
2260 &2807&	PG 1634+706	\\
\end{tabular}
\end{table}

\begin{table}[]
\caption[]{
Probability of observed excess in first bin as determined
from Monte Carlo simulations.  The strength of the correlation
function on the scale of 250-500 km/s, $\xi$, is large.
The 90\% confidence intervals, $\xi_{\rm min}$  and $\xi_{\rm max}$,
are determined by Monte Carlo simulations as
described in \S~IV.
They are conservative intervals in that they are
slightly $>90$\% confidence intervals.
\label{probtab}}
\begin{tabular}{cccccc}
Prob  &   Metal lines & $\Delta v $ or $\Delta r$ &
$\xi$ & $\xi_{\rm min}$  &  $\xi_{\rm max}$ \\
\hline

0.0047   &     no     &     $\Delta v$  & 1.8 & 0.6 & 3.4 \\
0.0011   &     yes    &     $\Delta v$  & 2.1 & 0.9 & 3.5 \\
0.0013   &     no     &     $\Delta r$  & 2.0 & 0.9 & 3.7 \\
0.0004   &     yes    &     $\Delta r$  & 2.2 & 1.0 & 3.5 \\

\end{tabular}
\end{table}

\begin{table}[]
\caption[]{Ly-$\alpha$ line pairs in the range 250-500 km/s.
Pairs with a dagger ($^\dagger$) are those pairs which include a
metal-line system.
Equivalent widths are given in rest frame coordinates.
All values are given in Angstroms.
\label{pairtab}}
\begin{tabular}{ccccl}
$\lambda$
& $\Delta \lambda$ &
EqW$_1$ &
EqW$_2$ &
Object  \\
\hline

   1485.8$^\dagger$&   1.6&  0.92&  0.34&   3C 351        \\
   1562.7&   2.4&  0.62&  0.37&   3C 351			\\        
   1364.4&   2.0&  0.32&  0.53&   H 1821+643                    \\
   2030.6&   2.9&  0.60&  0.40&   TON 153                       \\
   2033.3&   2.4&  0.39&  0.78&   TON 153                       \\
   2035.7&   2.5&  0.78&  0.78&   TON 153                       \\
   2037.9&   1.8&  0.78&  0.27&   TON 153                       \\
   1847.8&   2.4&  0.45&  0.25&   PG 1352+011     \\
   1850.5&   3.0&  0.25&  0.41&   PG 1352+011         \\
   1853.5$^\dagger$&   3.1&  0.41&  2.62&   PG 1352+011    \\
   2020.5&   3.1&  0.52&  0.34&   PG 1352+011                    \\
   2418.5$^\dagger$&   3.3&  0.25&  1.10&   PG 1634+706 \\
   2562.9&   4.0&  0.37&  0.28&   PG 1634+706   \\
   2601.6&   3.5&  0.90&  0.48&   PG 1634+706  \\

\end{tabular}
\end{table}




\begin{figure}[h]
 \caption{Observed number of pairs as a function of velocity separation
(solid lines) show a strong excess at the smallest velocity scales. 
The expected number of pairs in absence of clustering (dashed lines)
are given for two values of the number density evolution,
$\gamma = 0 ~{\rm and}~ 1.5$
as determined from extensive Monte Carlo simulations.
In the top panel, metal-line systems are excluded from the sample.}
\plotone{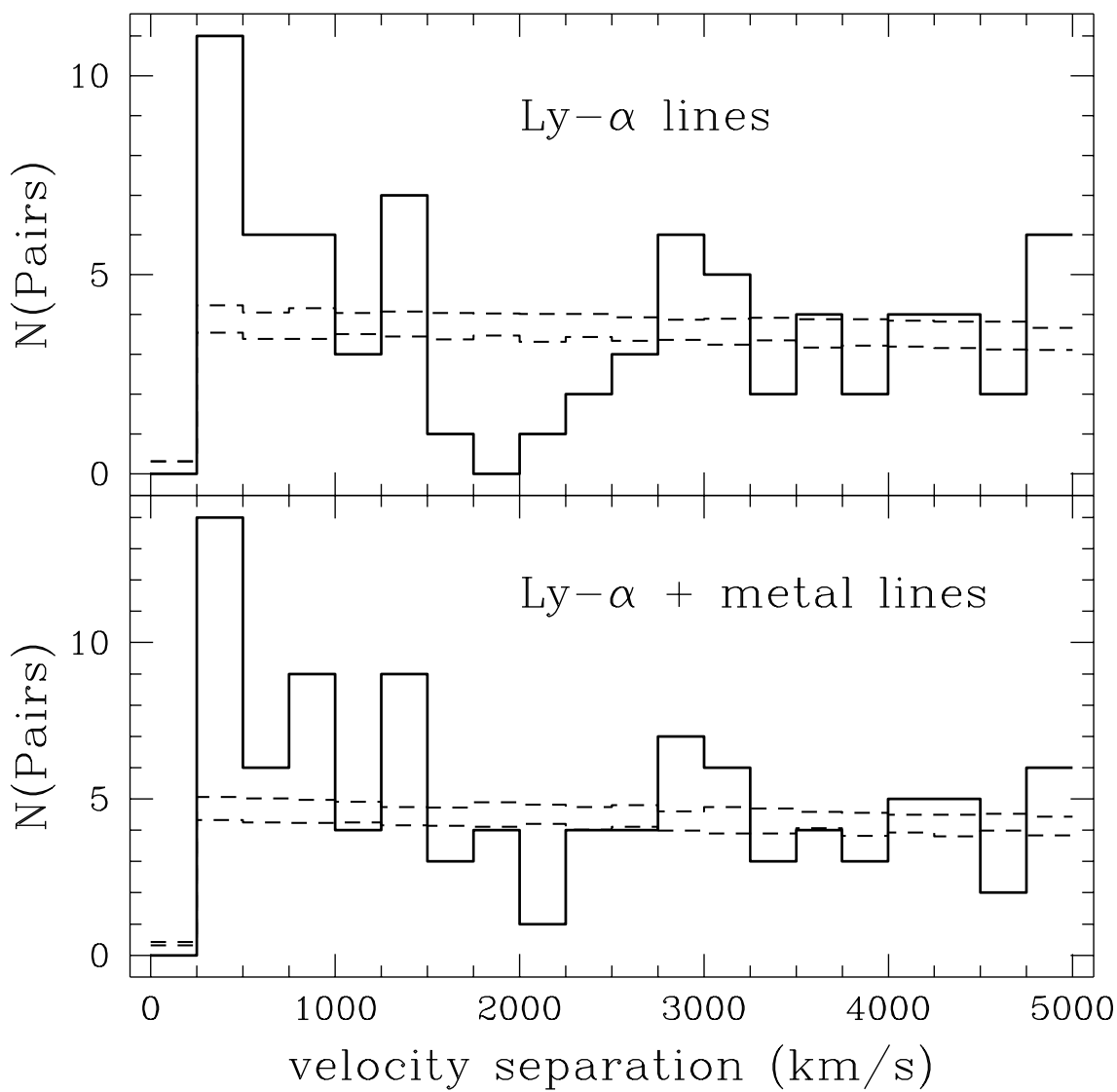}
\end{figure}

\begin{figure}[h]
 \caption{Same as Figure~1, except the number of pairs are measured as a
function of comoving separation assuming no peculiar velocities for the
absorbers.}
\plotone{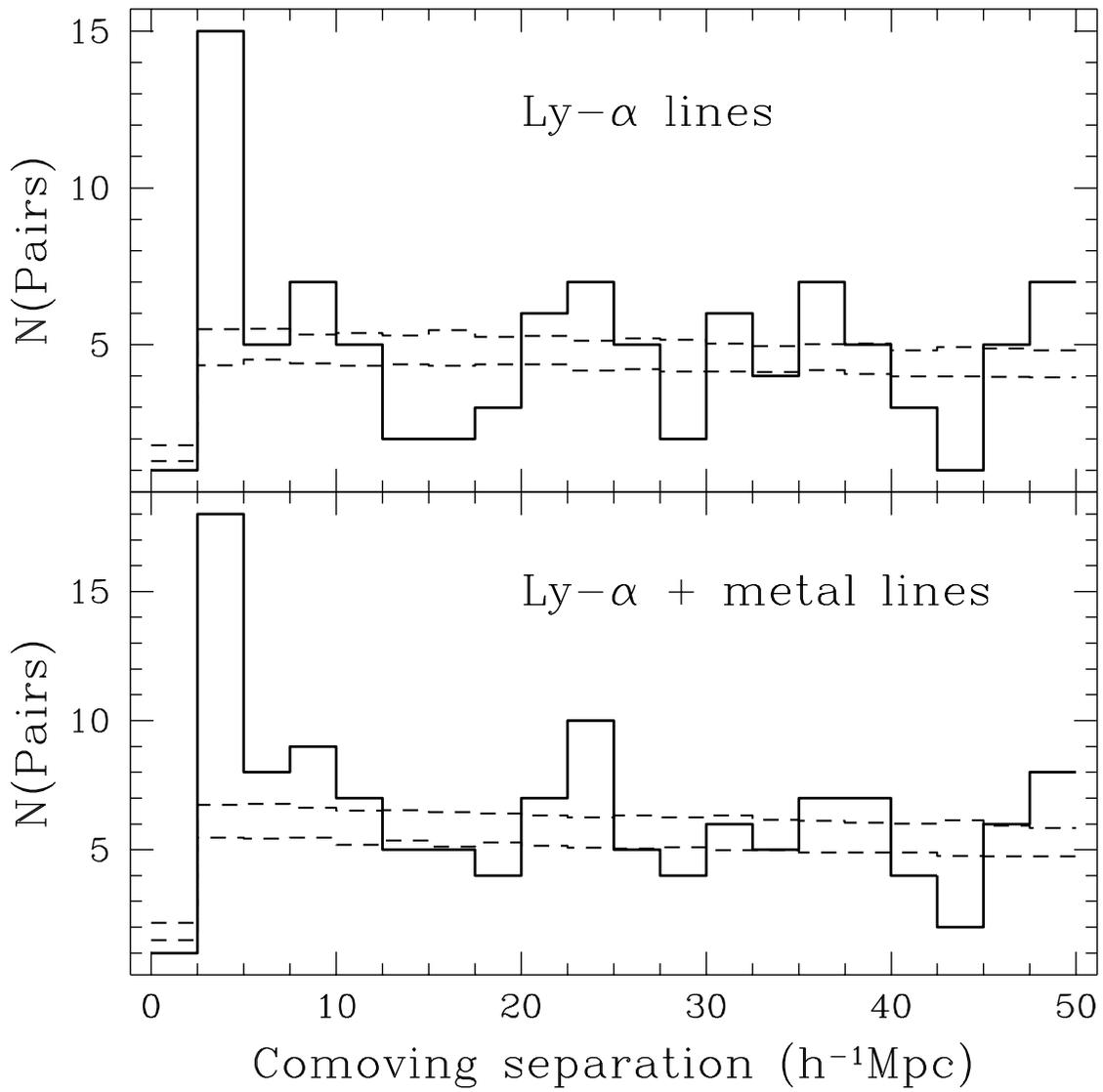}
\end{figure}

\begin{figure}[h]
 \caption{The top panel shows the observed number of pairs (solid line)
for the Ly-$\alpha$
lines (with metal lines excluded).
The upper and lower confidence intervals as determined by Monte Carlo
experiments are indicated with dashed lines.
As described in \S 4, confidence limits apply to the first bin alone.
The bottom panel shows the same observed number of pairs
(solid line) as in the top panel with the galaxy-galaxy correlation
function ($z=0$)
shown as a dashed line. The two functions are similar in strength;
however, the absorbers likely have peculiar velocities,
so the two correlation functions are not directly comparable.}
\plotone{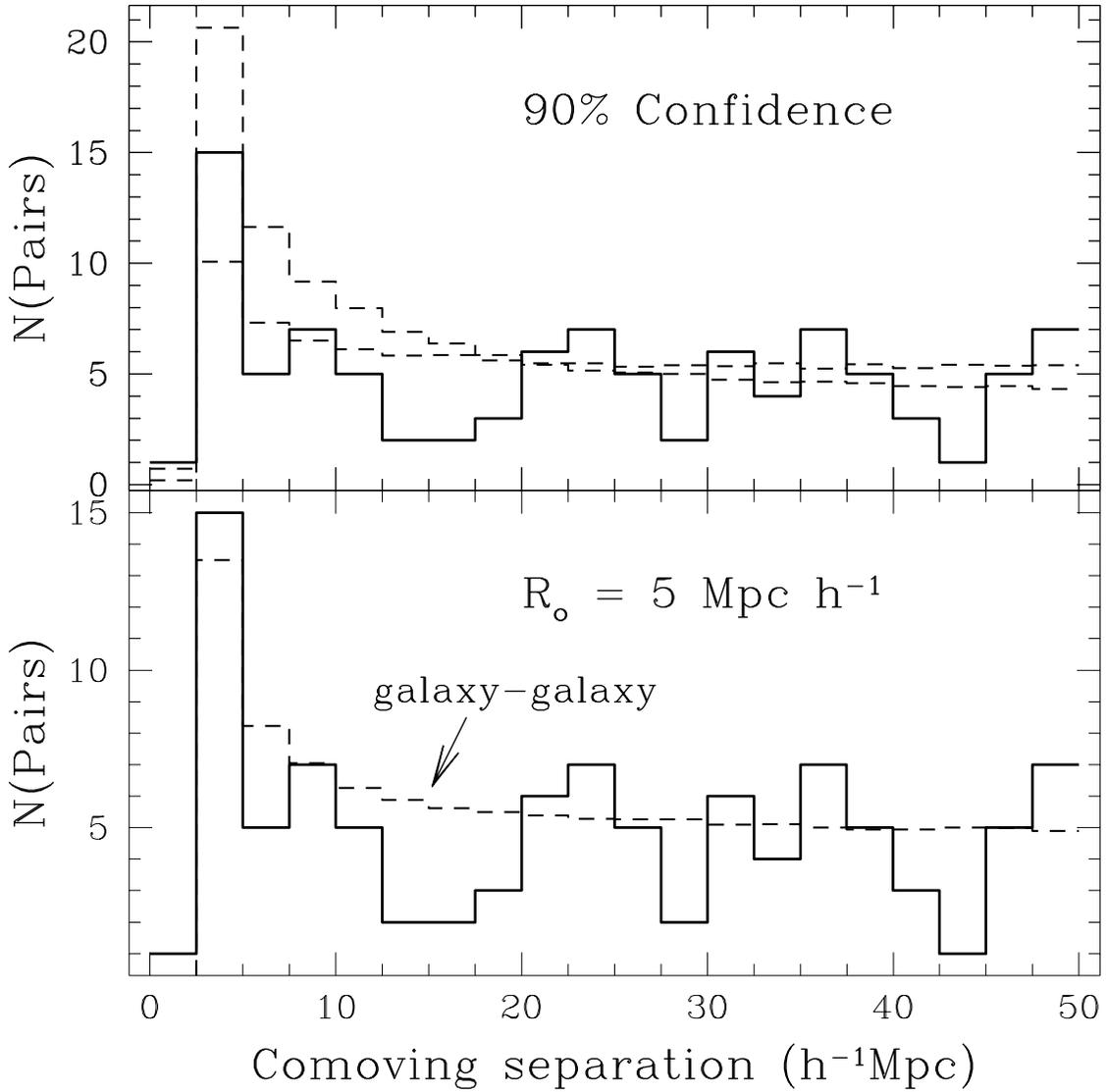}
\end{figure}

\end{document}